\documentclass{aa}

\usepackage{graphicx}
\usepackage{natbib}
\usepackage{txfonts} 
\begin{document}
  \title{Coronal activity cycles in nearby G and K stars}
  \subtitle{XMM-Newton monitoring of 61 Cygni and $\alpha$ Centauri}

   \author{J. Robrade\inst{1}
          \and
         J.H.M.M. Schmitt\inst{1}
         \and 
         F. Favata\inst{2}
          }


        \institute{Hamburger Sternwarte, Universit\"at Hamburg, Gojenbergsweg 112, 21029 Hamburg, Germany
              \and
        European Space Agency, 8-10 rue Mario Nikis, 75738 Paris Cedex 15, France\\
       \email{jrobrade@hs.uni-hamburg.de}
             }

   \date{Received; Accepted}

    \abstract
 {The existence of stellar analogues to the solar activity cycle is well established for chromospheric activity; in contrast
the investigation of coronal counterparts is just at its beginning.}
{An ongoing X-ray monitoring program of solar-like stars with XMM-Newton is performed 
to investigate coronal activity cycles.
   }
{We use X-ray observations of the nearby binaries 61 Cyg A/B (K5V and K7V) and $\alpha$\,Cen~A/B (G2V and K1V)
to study the long-term evolution of magnetic activity in weakly to moderately active G + K dwarfs over nearly a decade.
Specifically we search for X-ray activity cycles and related coronal changes and compare them to the solar behavior.
   }
  {For 61\,Cyg~A we find a regular coronal activity cycle analog to its 7.3~yr chromospheric cycle.
The X-ray brightness variations are with a factor of three significantly lower than on the Sun, 
yet the changes of coronal properties resemble the solar behavior with larger variations occurring in the respective hotter plasma components.
61\,Cyg~B does not show a clear cyclic coronal trend so far, but the X-ray data matches the more irregular chromospheric cycle.
Both $\alpha$\,Cen stars exhibit significant long-term X-ray variability.
$\alpha$\,Cen~A shows indications for cyclic variability of an order of magnitude with a period of about 12\,--\,15 years; 
the $\alpha$\,Cen~B data suggests an X-ray cycle with an amplitude of about six to eight and a period of 8\,--\,9 years.
The sample stars exhibit X-ray luminosities ranging between $L_{\rm X} \lesssim 1 \times 10^{26} - 3 \times 10^{27}$~erg\,s$^{-1}$
in the 0.2\,--\,2.0~keV band and have coronae dominated by cool plasma with variable average temperatures of around 1.0\,--\,2.5~MK.
}
{Coronal activity cycles are apparently a common phenomenon in older, slowly rotating G and K stars.
The spectral changes of the coronal X-ray emission over the cycles are solar-like in all studied targets.
}


   \maketitle
%

\section{Introduction}

The 11 years activity cycle of the Sun is one of the key characteristics of solar magnetic activity.
The most evident feature of the solar cycle and those with the longest recorded history
is the periodic modulation of sunspot number. Other activity indicators like chromospheric \ion{Ca}{ii} emission or coronal X-rays follow the sunspot cycle, but
magnetic field studies have shown that the true cycle period
is actually 22 years due to the polarity reversal of the solar magnetic field.
The solar cycle is not exactly periodic and both its amplitude and length vary; even periods without significant activity like
the 17th century Maunder Minimum have been observed. 
Solar activity also affects the Earth's environment and apparently at least local climate conditions
through modulations of the magnetosphere, high energy particle flux or irradiance.

Studies of multiple activity indicators reveal that the amplitude of cyclic
variation strongly depends on the indicator used.
X-ray studies of the Sun show a pronounced coronal cycle
with about one order of magnitude brightness variations between solar maximum and minimum, e.g. 
$\log L_{\rm X} \approx 26.5 \dots 27.5$~erg\,$s^{-1}$ in the ROSAT 0.1\,--\,2.4~keV energy band \citep{per00}.
However, for the Sun the calculation of a disk integrated X-ray flux like it is observed for stars is far from trivial, 
as can be concluded from the divergent X-ray brightness estimations presented by \citet{jud03}.
While at harder X-ray energies the variations can easily be of the order of hundred as observed by the {\it Yohkoh} satellite \citep{orl01}, 
in the chromospheric \ion{Ca}{ii}~H+K lines or in transition region lines observed at UV wavelengths these much lower ranging from several tens of percent up to a factor of a few
\citep[e.g.][]{lea97}. 

The solar behavior raises the question, if activity cycles also exist on other stars.
These studies naturally require long-term monitoring and the Mt. Wilson program of \ion{Ca}{ii}~H+K emission was the first to show that chromospheric
activity cycles exist on many late-type stars \citep{bal95}. Indeed, chromospheric activity cycles with periods of $P_{\rm cyc} \approx 7 \dots 15$~yr are rather common on slowly rotating and thus weakly to
moderately active G0\,--\,K5\,dwarf stars. A few of the less active stars show flat activity curves and likely reside currently in a 
Maunder Minimum state, with the G2 star 51~Peg being an example that was also observed in X-rays \citep{pop09}.
In contrast, the more active stars do not show cyclic but rather erratic chromospheric variability.

Our {\it XMM-Newton} monitoring program targets the long-term evolution of stellar X-ray emission and we have now 
data for almost a decade that
enables us to address the question of the existence of stellar X-ray cycles and allows to study the cycle properties 
for a number of individual stars. 
The program focusses on a few nearby solar-type dwarf stars with spectral types from early G to mid K, corresponding to
a masses of $1.1 \dots 0.7$~M$_{\odot}$ and $T_{\rm eff} \approx 4200 \dots 5800$~K. 
All stars are weakly to moderately active in X-rays, i.e. $\log L_{\rm X}/L_{\rm bol} \approx -5.5 \dots  -7$
and three out of five were known to exhibit a chromospheric cycle.
Beside the wide binaries 61\,Cyg A/B and $\alpha$\,Cen~A/B that are described in detail below, 
the monitoring sample includes the G2 star HD~81809 \citep{fav08}.

In this paper we present results from the {\it XMM-Newton} monitoring of 61\,Cyg A/B and $\alpha$\,Cen~A/B.
Our paper is structured as follows; in Sect.\,\ref{tar} we present the target stars, in Sect.\,\ref{obsana} we describe observations and data analysis,
in Sect.\,\ref{res} we discuss X-ray light curves and short-term variability, in Sect.\,\ref{cyc}
we investigate coronal activity cycles and compare our results to solar measurements 
and we close with a summary in Sect.\,\ref{summ}.

\section{The target stars} 
\label{tar}

\subsection{61 Cygni}

The 61\,Cyg~A/B (\object{HD 201091}, \object{HD 201092}) system is a nearby visual K dwarf binary at a distance of 3.5~pc, consisting of a K5V and K7V star. 
Both stars are slow rotators with periods of 35 and 38 days \citep{don96} and the system is thought to be several Gyr old;
age estimates range from 2\,--\,3~Gyr \citep{bar07} up to $6 \pm 1.7$~Gyr \citep{eps12}.
Interferometric radii measurements are presented in \citet{ker08}, combined with evolutionary models they indicate an age of $6 \pm 1$~Gyr.
The stars 61\,Cyg A and B are included in the Mt. Wilson \ion{Ca}{ii}~H+K program and
both stellar components were found to be cyclic with respective periods of 7.3~yr (A) and 11.7~yr (B) \citep{bal95}. 
The chromospheric cycle of the 61\,Cyg~A (mean S-index of $\langle S \rangle = 0.66$) is quite regular, 
while the chromospherically more active component
61\,Cyg~B ($\langle S \rangle = 0.99$) has a less smooth cycle and shows more irregular variations.
Like the Sun, 61\,Cyg~A has with an 'excellent' cycle the highest quality grade based on false alarm probability used by \citet{bal95}.
Further it is also the reddest (largest B-V) and most active (largest  $\langle S \rangle$) star with this grade in the complete sample of over hundred stars.
A follow-up program was started at Lowell Observatory that continues to observe 61\,Cyg in chromospheric emission lines;
these more recent data are presented in \citet{hall07}.

A {\it ROSAT} X-ray monitoring program of 61\,Cyg tracked the X-ray behavior of the moderately active components over a few years in the 1990s,
$L_{\rm X} = 1 \dots 3 \times 10^{27}$~erg\,s$^{-1}$ (A) and $L_{\rm X} = 0.4 \dots 1\times 10^{27}$~erg\,s$^{-1}$ (B),
but the time basis of these observations was too short to cover a complete cycle \citep{hem03}. 
The first years of {\it XMM-Newton} X-ray monitoring of 61\,Cyg are discussed in \citet{hem06} and
an update is given in \citet{rob07cy}. The data again covered only half of the cycle period, but already  
showed that a coronal cycle in phase with the chromospheric cycle exists on 61\,Cyg~A. 
An analysis of the merged RGS spectra of 61\,Cyg and $\alpha$\,Cen focussing on neon and oxygen lines in weakly active stars is presented in \citet{rob08}.

\subsection{$\alpha$ Centauri}
The visual binary $\alpha$\,Centauri~A/B (\object{HD 128620}, \object{HD 128621})
consists of a G2V (A) and a K1V (B) star and is part of the nearest stellar system at a distance of 1.35\,pc; it also includes
the M5~dwarf Proxima Centauri which is outside the FOV of our observations.
The two components $\alpha$\,Cen A and B are separated by about 23~AU and have an orbital period of roughly 80 years.
Basic stellar properties are e.g. presented by \citet{fla78} and interferometric diameters are determined by \citet{ker03}.
The $\alpha$\,Cen system is thought to be slightly older than the Sun with age estimates ranging from $4.2-5.2$~Gyr \citep{bar07} to $6.8 \pm 0.5$~Gyr \citep{eps12}.
Both stars are slow rotators with periods of around 29\,(A) and 37\,(B) days and have weakly active coronae as
shown by X-ray observation obtained in 1979 with the {\it Einstein} observatory. 
The secondary $\alpha$\,Cen~B was with $L_{\rm X} = 2.8 \times 10^{27}$~erg\,s$^{-1}$ compared to $L_{\rm X} = 1.2 \times 10^{27}$~erg\,s$^{-1}$ for $\alpha$\,Cen~A the 
X-ray brighter component as shown by \citet{gol82}, who brought up the idea that both stars might have activity cycles that are out of phase by 
comparing the {\it Einstein} data with previous {\it IUE} (International Ultraviolet Explorer) observations.
$\alpha$\,Cen~B was again the X-ray brighter component in {\it ROSAT} observations in the 1990s \citep{schmitt97}.
The first grating spectra obtained in 1999 with the {\it Chandra} LETGS, providing wavelength coverage up to 175~\AA,
show that the A/B flux ratio is temperature, respectively energy band dependent \citep{raa03}.
With average coronal temperatures around 1.6~MK (A) and 1.8~MK (B)
$\alpha$\,Cen~B was the X-ray brighter component above 0.1~keV,
but in softer X-rays they approach equal luminosity
and $\alpha$\,Cen~A was even brighter in emission lines originating from very cool plasma at $\lesssim 1$~MK.
Other similarities of the $\alpha$\,Cen stars to the Sun were discovered, e.g. elemental abundance pattern that show the FIP-effect (First Ionization Potential), 
i.e. an coronal enhancement of low FIP elements compared to high FIP elements.

For the $\alpha$\,Cen system no chromospheric long-term monitoring data are available 
and both stars were not known to exhibit activity cycles when we started our observations.
Early results of the {\it XMM-Newton} monitoring program of the $\alpha$\,Cen stars are presented by \citet{rob05}.
Most strikingly an X-ray dimming of $\alpha$\,Cen~A by an order of magnitude within three years was observed in the bandpass of the EPIC detectors,
caused by a strong decrease in emission measure at a few MK.
Shortly afterwards the secondary $\alpha$\,Cen~B
also started to become fainter in X-rays and indications for cyclic behavior emerged when
observations verified the ongoing decline of its X-ray brightness \citep{rob07cy}. 
The $\alpha$\,Cen system is also monitored with the {\it Chandra} HRC-I since the
end of 2005. The HRC is quite sensitive to soft X-ray photons and provides superb spatial resolution and positions, but virtually no spectral information.
A second LETGS observations from 2007 put the analogy of the $\alpha$\,Cen~A luminosity decrease
to the solar behavior, i.e. a vanishing of predominantly hotter coronal structures and a rather constant flux from cooler ($\lesssim$~1MK) structures, 
on a firmer ground \citep{ayr09}. Average coronal temperatures in 2007 were 1.1~MK for $\alpha$\,Cen~A and 1.9~MK for $\alpha$\,Cen~B.
Extensive UV coverage of the $\alpha$\,Cen system is provided e.g. by {\it IUE} 
and {\it FUSE} ({\it Far Ultraviolet Spectroscopic Explorer}).
In a multi-wavelength study of $\alpha$\,Cen~B, \citet{dew10} analyze measurements at X-ray, FUV and UV wavelengths 
and determine an activity cycle period of $8.8 \pm 0.4$~yr.
A similar period of $8.4$~yr was derived by \citet{buc08} utilizing {\it IUE} \ion{Mg}{ii} line fluxes and \ion{Ca}{ii} measurements
including new data obtained at the CASLEO observatory.
However, no clear activity cycle for $\alpha$\,Cen~A has been detected so far.

\section{Data analysis and observations}
\label{obsana}

\subsection{Overview}
The target 61\,Cyg has been observed twice a year with {\it XMM-Newton} since 2002,
while $\alpha$\,Cen has been observed with a  more irregular sampling since 2003.
We present an analysis of the complete data obtained until the end of 2011 and we study
20 exposures for 61\,Cyg and 12 exposures for $\alpha$\,Cen with
individual exposure times ranging between about 5\,--\,16~ks. In total 61\,Cyg was observed for about 225~ks and $\alpha$\,Cen for 115~ks;
observations used in this work are summarized in Table~\ref{allobs}.

The data were reduced with the {\it XMM-Newton} Science Analysis System
software SAS~10.0, older data was reprocessed and standard SAS tools \citep{sas} were used
to produce images, light curves and spectra.
For our analysis we use data from the EPIC (European Photon Imaging Camera) instrument that consists of two MOS and one PN detector;
importantly the point spread function (PSF; FWHM $\sim 6\,\arcsec$) is better sampled by the MOS (pixel size $1.1\,\arcsec$), while the PN has a higher sensitivity.
A detailed description of the instruments can be found in the 'XMM-Newton Users Handbook' \footnote{(http://xmm.esac.esa.int).}
Spectral analysis was carried out with XSPEC V12.6 \citep{xspec} and we use multi-temperature APEC/VAPEC models \citep{apec} 
with abundances relative to solar values as given by \citet{grsa} to derive X-ray luminosities and emission measure distributions (EMD). 
Source and background photon extraction is adjusted to the individual source properties as described below.
The spectral analysis is performed independently for MOS and PN data when possible and we obtain overall consistent results, but
discrepancies of about 20\,\% are present for individual datasets. The quoted errors are statistical errors at a 90\,\% confidence level;
note that the intrinsic X-ray brightness variability of the target stars is typically larger than any statistical or systematic error
except for the very faintest observations.
For the study of activity cycles we investigate the quasi-quiescent coronal state, thus
we exclude larger flares on the basis of the respective X-ray light curves; selected flares are discussed in a separate section.
Here strong flaring activity is defined as phases where the peak flux is above the average quasi-quiescent level by a factor of two or more.
All given quasi-quiescent properties refer to the 0.2\,--\,2.0~keV energy band.
To convert our 0.2\,--\,2.0~keV luminosities to typical energy bands
of {\it Einstein} (0.15\,--\,4.0~keV) and {\it ROSAT} (0.1\,--\,2.4~keV) and vice versa
we calculate temperature dependent conversion factors from APEC models. 
In the temperature range 2\,--\,10~MK the differences in energy band flux are relatively minor
compared to the intrinsic variability. For example at 2~MK we obtain 1.1/1.15 times higher fluxes in the respectively softer bands, but
towards lower temperatures the factors increase, e.g. they are 1.5/1.9 at 1~MK (see Table~\ref{apecc}).

Determination of surface fluxes $F_{\rm X}$ and activity levels $L_{\rm X}/L_{\rm bol}$ adopt the following radii and luminosities taken 
from above literature or calculated from the visual magnitudes: 1.2~$R_{\odot}$, 1.5~$L_{\odot}$
and 0.9~$R_{\odot}$, 0.5~$L_{\odot}$ for $\alpha$\,Cen A/B; as well as 0.7~$R_{\odot}$, 0.15~$L_{\odot}$ and 0.6~$R_{\odot}$, 0.1~$L_{\odot}$ for 61\,Cyg~A/B. 

\begin{figure}[t]
\begin{center}
\includegraphics[width=85mm]{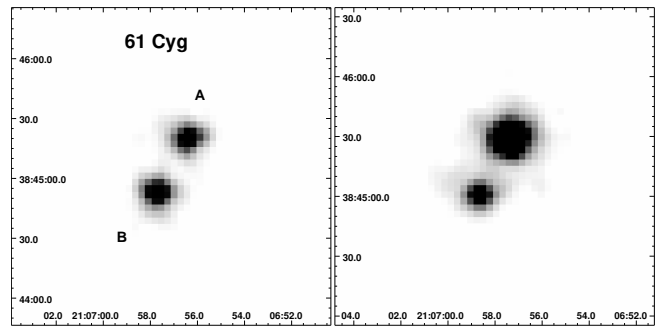}
\end{center}
\caption{\label{pics61}MOS1 images of 61\,Cyg taken in 2006 (left) and 2009 (right) around activity minimum and maximum for 61\,Cyg~A.
Exposure times are similar and an identical linear brightness scaling is used.}
\end{figure}

\begin{figure}[t]
\begin{center}
\includegraphics[width=80mm]{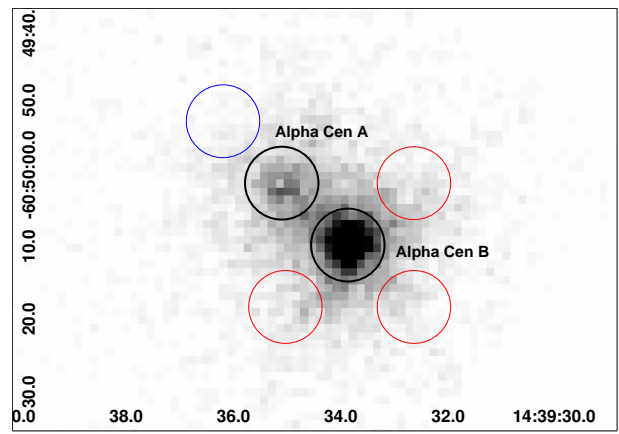}
\end{center}
\caption{\label{picsac}MOS1 image of $\alpha$\,Centauri A/B obtained in 2003, source regions (black) and control regions (red/blue) are overplotted.}
\end{figure}

\subsection{Modeling of 61 Cygni and $\alpha$ Centauri}
\begin{figure}[t]
\begin{center}
\includegraphics[width=90mm]{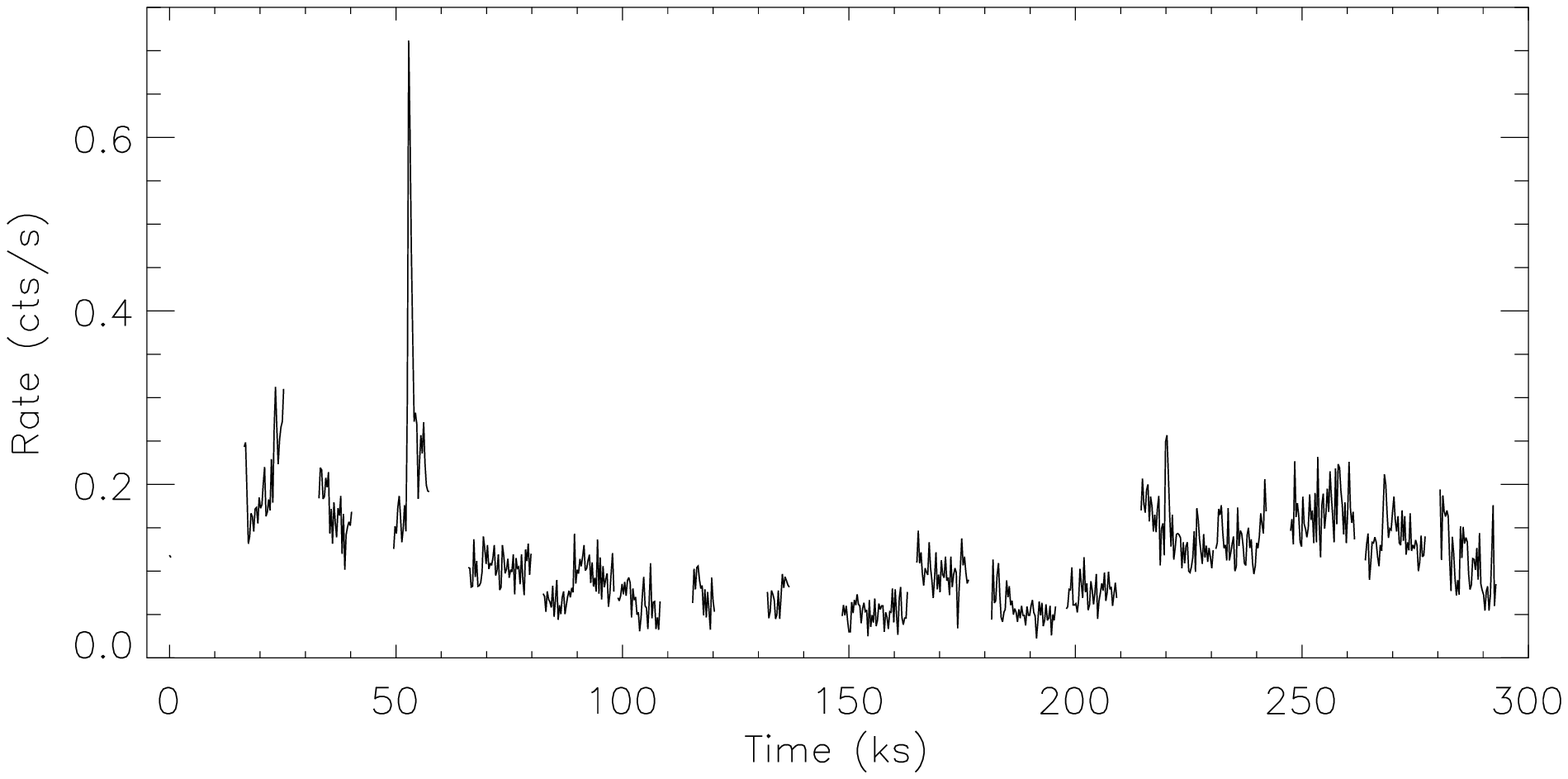}

\smallskip
\includegraphics[width=90mm]{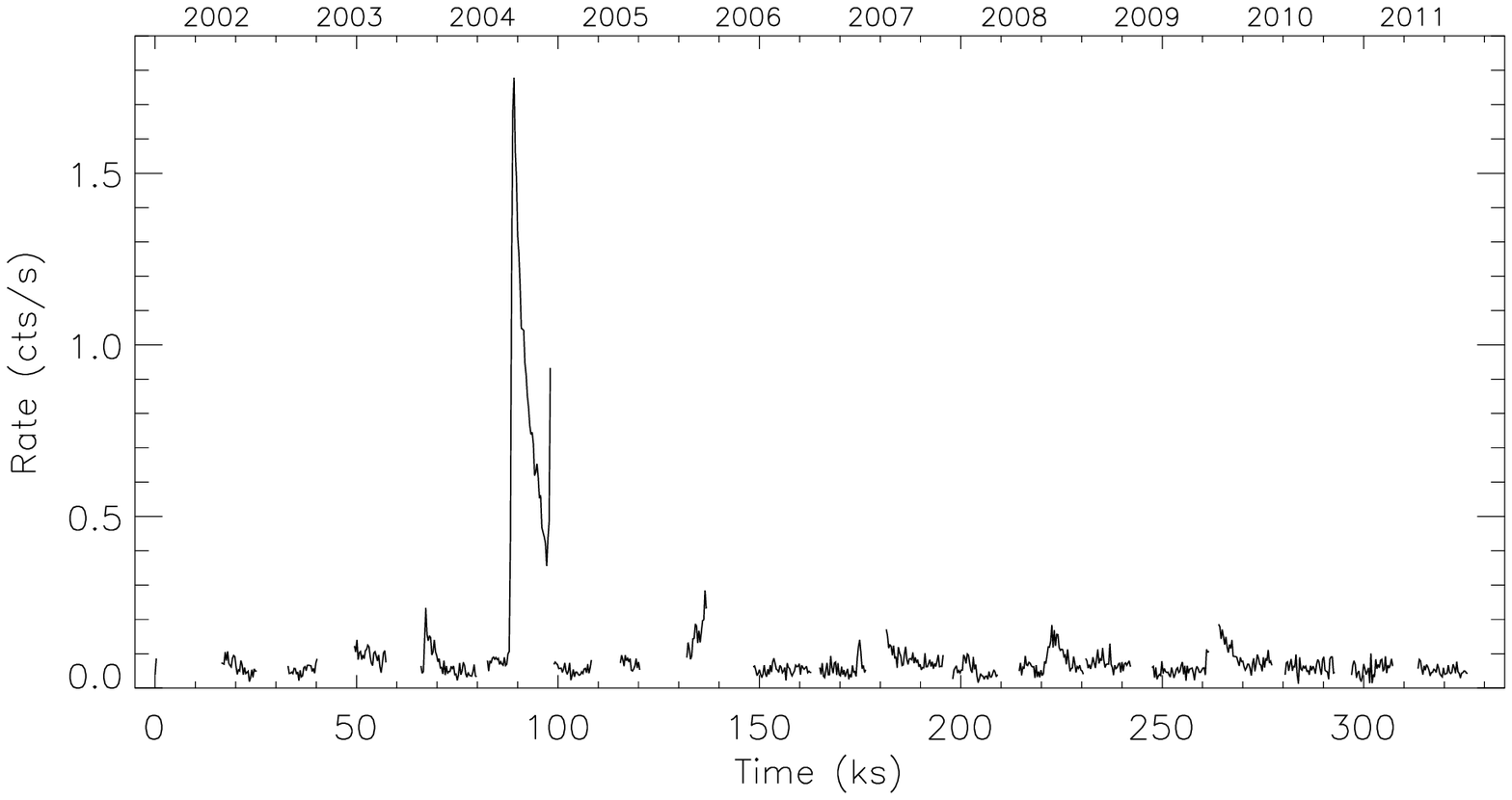}
\end{center}
\caption{\label{61lcs}X-ray light curves of 61 Cyg A (top) and 61 Cyg B (bottom) during the XMM monitoring, merged MOS data.}
\end{figure}

The two components of 61\,Cyg A/B are with a distance of roughly $33\,\arcsec$ well separated in all EPIC detectors as shown in Fig.\,\ref{pics61}.
Source photons were extracted from a $15\,\arcsec$ circular region around the respective component and the background was taken from nearby source free regions.
The contamination from the other component is less than 3\,\% and was not removed except for very large flares (see Fig.\,\ref{61lcs}).
For the 05/2002 observation we use only the PN data for spectral analysis due to 
the very short ($< 1$~ks) exposure of the MOS detectors and
we correct the count rates of the 10/2010 and 04/2011 observations for vignetting due to offset pointings by using instrument effective area files.
We find that 3-T models are sufficient to describe the spectra and
applied two types of models that have temperatures as free parameters or use a fixed 0.1, 0.3 and 0.7~keV temperature grid.
Both models have solar metallicity and the derived results agree with each other.
We also tested models with lower (0.6 solar) or variable abundances, but again find only minor effects on the derived X-ray luminosities.

The distance between the $\alpha$\,Cen A/B binary components was about $12\,\arcsec$ in 2003 and it
declined throughout our monitoring to $7\,\arcsec$ in 2010.
Thus the binary is, especially later in the campaign, only moderately resolved by {\it XMM-Newton} and photon extraction becomes crucial.
We use the position of the X-ray brighter component $\alpha$\,Cen~B as zero-point of the system and determine those of $\alpha$\,Cen~A from
the respective offset. Source photons are extracted from circular regions with $30\,\arcsec$ radius for the combined system and
$5\,\arcsec$ for both components (2003\,--\,2006) that gradually decreases thereafter.
For the study of X-ray fainter $\alpha$\,Cen~A,
MOS2 data are discarded due to its triangular PSF shape and
the PN data are used only until 2007, when the binary separation became comparable to twice its pixel size.
We study the effects of component cross-talk by using three adjacent control regions for $\alpha$\,Cen~A and one for $\alpha$\,Cen~B as shown in Fig.\,\ref{picsac}. 
A secondary, mirrored control region (not plotted) is used
to track the contribution from $\alpha$\,Cen~B itself, compared to those from $\alpha$\,Cen~A.
In summary, the contamination of $\alpha$\,Cen~A is with $\sim 30\,\%$ moderate in the beginning of the campaign, 
but becomes with $\gtrsim 90\,\%$ dominant in several later observations, those
of $\alpha$\,Cen~B is with $\lesssim 5\,\%$ minor for all observations, whereas the 'true' background is negligible. 
Light curves account for dead CCD columns in individual MOS exposures
and count rates are scaled regarding reduced detector 'live time' in the small/large window mode (2002\,--\,2007) 
and the narrower extraction regions (2007\,--\,2010).

To determine the X-ray luminosities of $\alpha$\,Cen~A/B, we first transform the measured count rates directly into X-ray fluxes by using temperature dependent conversion factors 
as calculated with the webPIMMS tool \footnote{(http://heasarc.gsfc.nasa.gov/Tools/w3pimms.html)}; here
we use solar abundance APEC models with temperatures of 1.0, 1.2, 1.4, 1.6 and 1.8~MK for the different activity phases, 
i.e. count rates, of $\alpha$\,Cen~A and 1.8, 2.0 and 2.2~MK for $\alpha$\,Cen~B (see Table~\ref{apecc}).
In a second step we model the spectra of the system
and determine the individual fluxes from the corrected A/B count-ratio. In addition,
spectra from the respective PSF cores of the components are modeled.
The background for $\alpha$\,Cen~A is taken from the control regions, otherwise nearby source free regions are used.
Spectral modeling uses multi-temperature VAPEC models and depending on its quality we use one, two or three temperature components.
To account for the FIP-effect found in the coronae of the $\alpha$\,Cen stars \citep{raa03}
we use two sets of abundances (a) 1.5 solar for low FIP elements (Fe, Mg, Ca, Ni), 0.5 solar for high FIP elements (O, N, Ne) and solar abundances for other elements and (b) O, N, Ne, Mg, Fe as free parameter and the remaining elements at solar.
Altogether we apply up to three methods to derive the X-ray fluxes, i.e. count rate conversion, weighted spectral modeling of the system and individual spectral modeling,
that are averaged to derive the final X-ray luminosity.

\section{X-ray monitoring}
\label{res}

The X-ray light curves obtained over our campaign give an impression of the diversity in variability of our target stars;
beside long-term variations of the quasi-quiescent flux also several flares have occurred during our observations.

\subsection{The total light curves}

The X-ray light curves of 61\,Cyg A and B are shown in Fig.\,\ref{61lcs}.
Prominent features are the smooth variation of the quasi-quiescent level of 61\,Cyg~A, a large flare on 61\,Cyg~B and an intermediate flare on 61\,Cyg~A.
In addition, many smaller flares, minor activity and variable activity level were recorded during the campaign.
A few observations were performed during an overall more active phases and moderate
variability of up to $\sim 50$\,\% is present in several of the individual datasets. 
The shown X-ray light curves are derived from the merged MOS data in the 0.2\,--\,2.0~keV energy band with background subtraction and
300~s time binning. Further we roughly establish real time chronology of the long-term evolution
by using a time interval of 16.5~ks for each exposure when plotting the light curves: this is indicated by the time line on top of each panel
while the x-axis tracks the individual running time.  This approach 
also allows a better comparison with Fig.\,\ref{61cyga} and \ref{61cygb}.

\begin{figure}[t]
\begin{center}
\includegraphics[width=90mm]{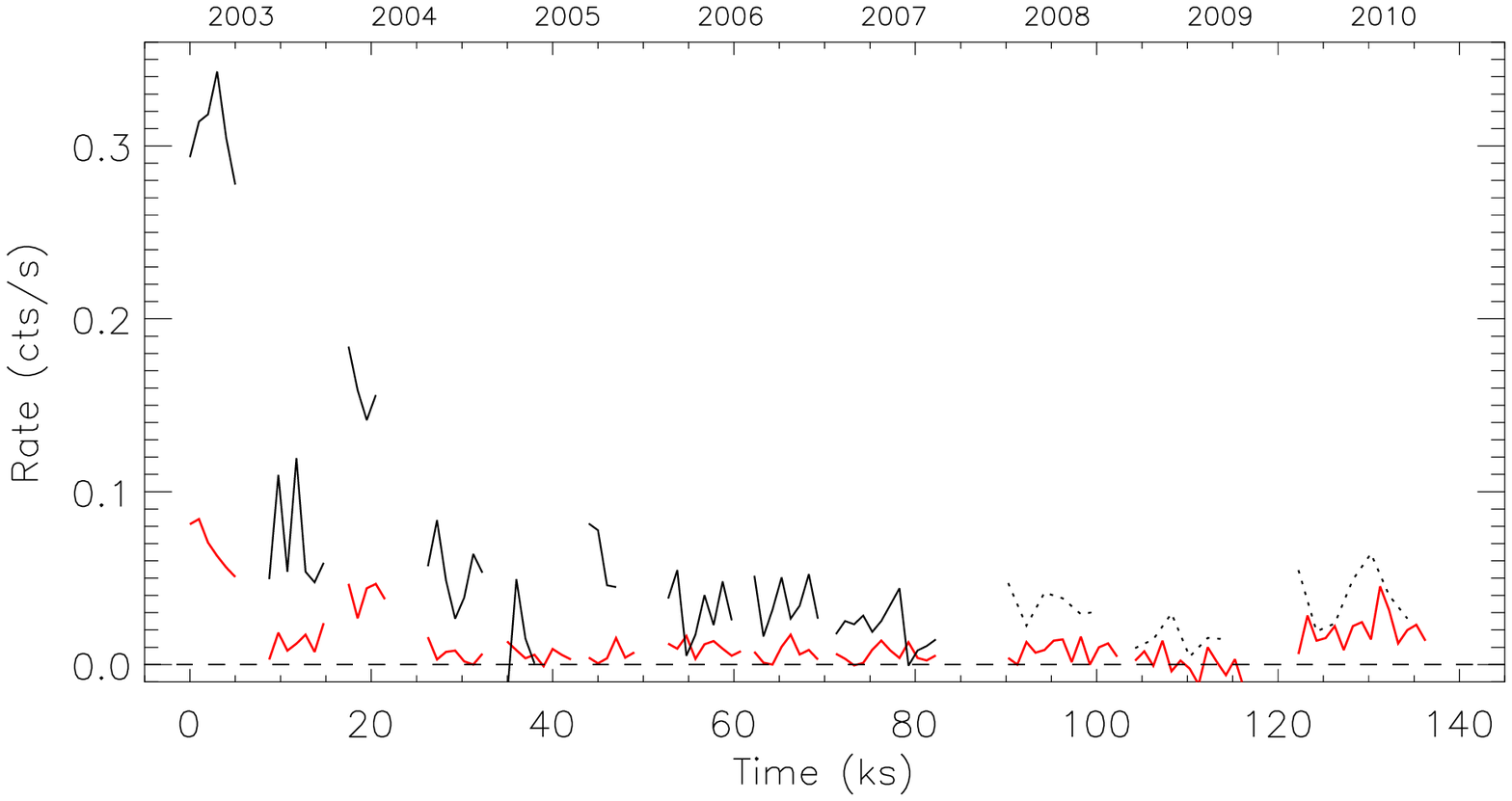}

\smallskip
\includegraphics[width=90mm]{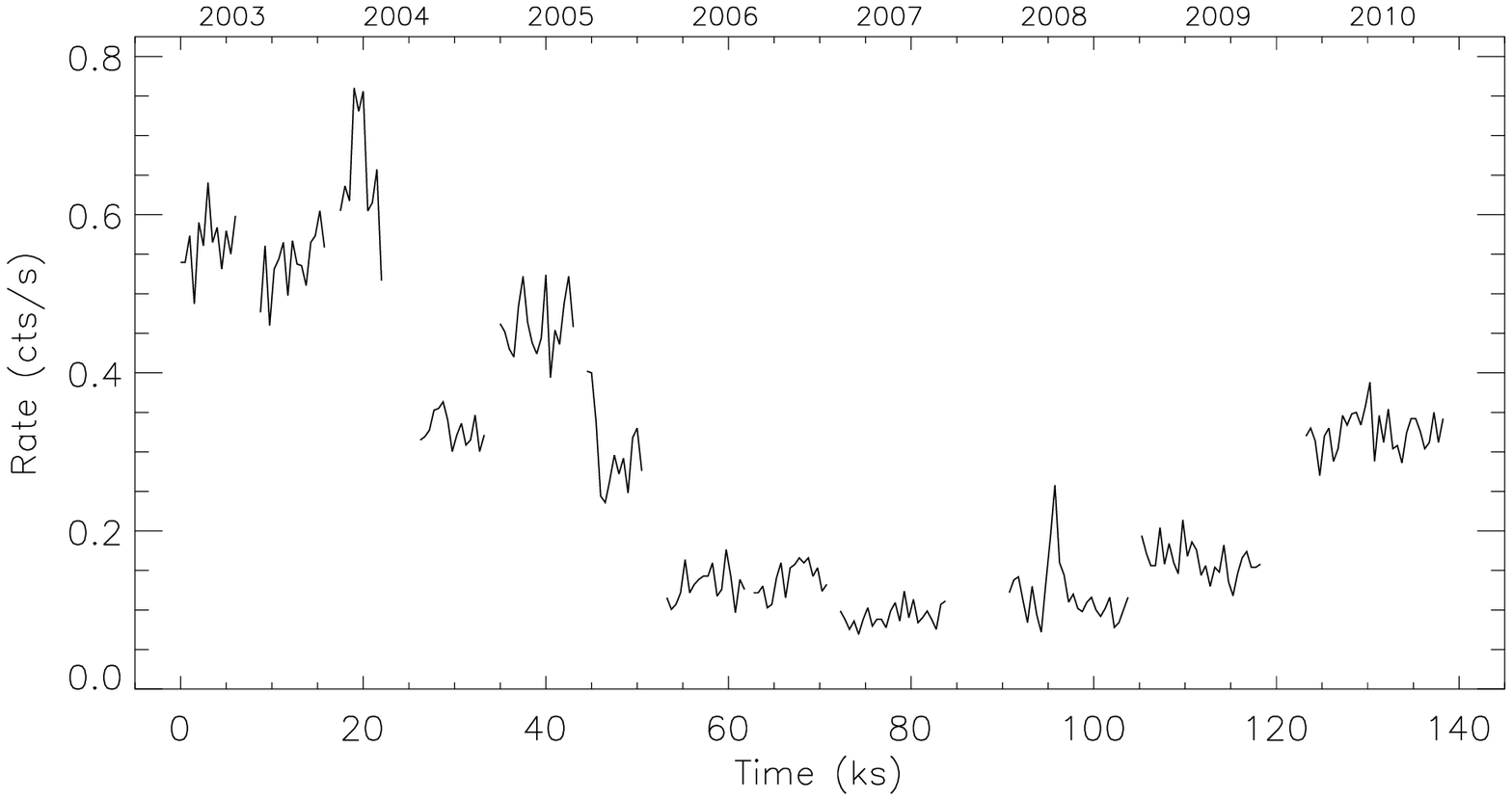}
\end{center}
\caption{\label{aclcs}{\it Top:} X-ray light curves of $\alpha$\,Cen~A from PN (black) and MOS1 (red). {\it Bottom:} $\alpha$\,Cen~B (merged MOS) during the XMM monitoring.}
\end{figure}

The light curves of $\alpha$\,Cen A/B are shown in Fig.\,\ref{aclcs};
for $\alpha$\,Cen~B we use merged MOS data with a binning of 500~s,
light curves of $\alpha$\,Cen~A were obtained independently from MOS1 and PN data with a binning of 1~ks, PN data excluded from analysis are indicated by dashed lines.
We use the 0.2\,--\,2.0~keV energy band and
approximate real time chronology of the long-term evolution by using 18~ks for each year in the plot.
Both components show significant long-term variability during our campaign.
They are characterized by low activity, in addition we detect two smaller flares on $\alpha$\,Cen~B, but no clear flare event from $\alpha$\,Cen~A.
The observed short-term variability of the $\alpha$\,Cen system stars is quite small,
but the low number of detected counts prevents a detailed study for most datasets.

\subsection{Flares}

We have detected several flares from each of our low to moderately active stars with the exception of the least active star $\alpha$\,Cen~A.
While not as strong and frequent as on more active stars,
the observed flares cover several orders of magnitude in maximum X-ray brightness and emitted energy.

\begin{figure}[t]
\centering
\includegraphics[width=80mm]{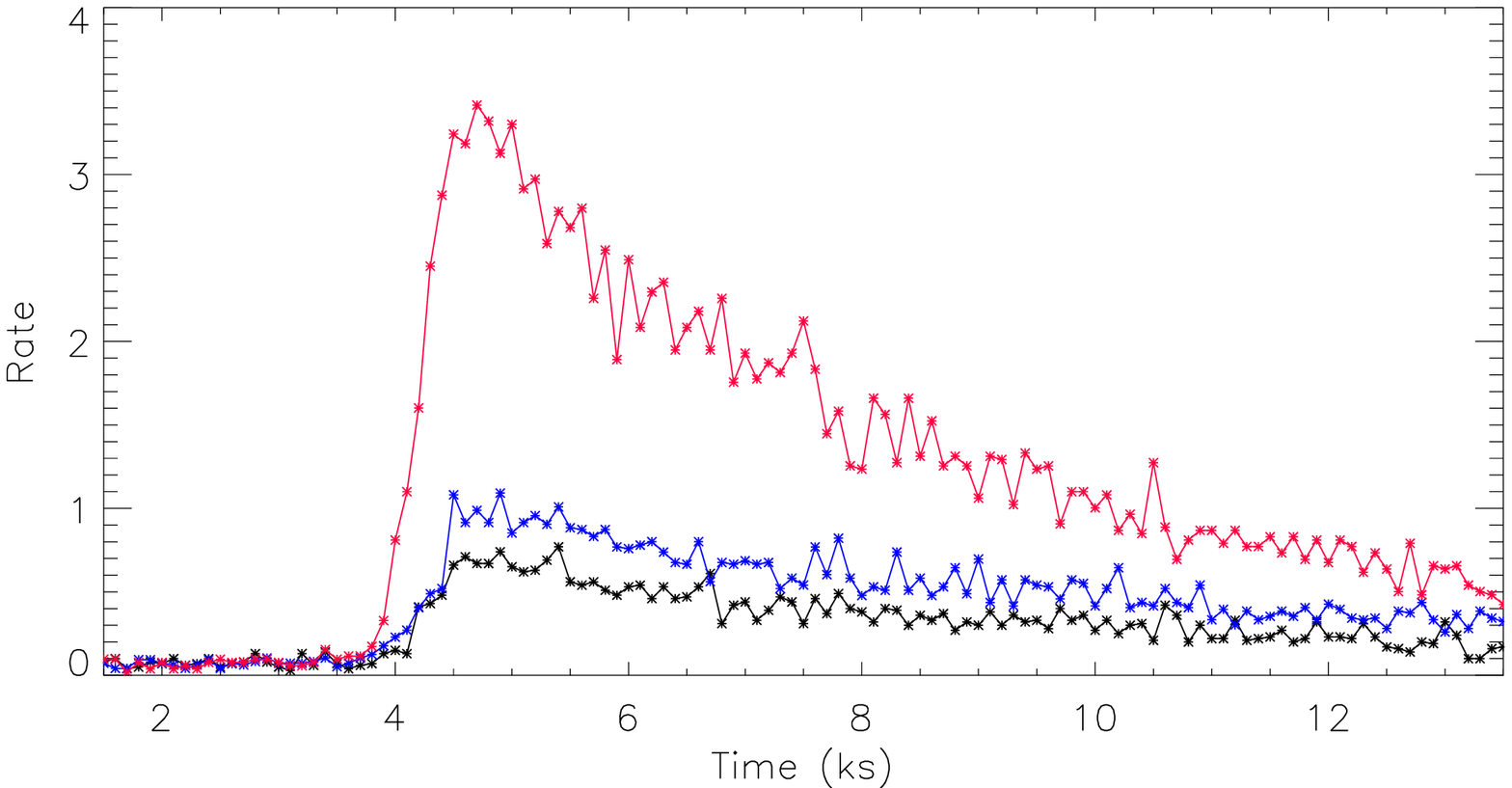}
\includegraphics[width=80mm]{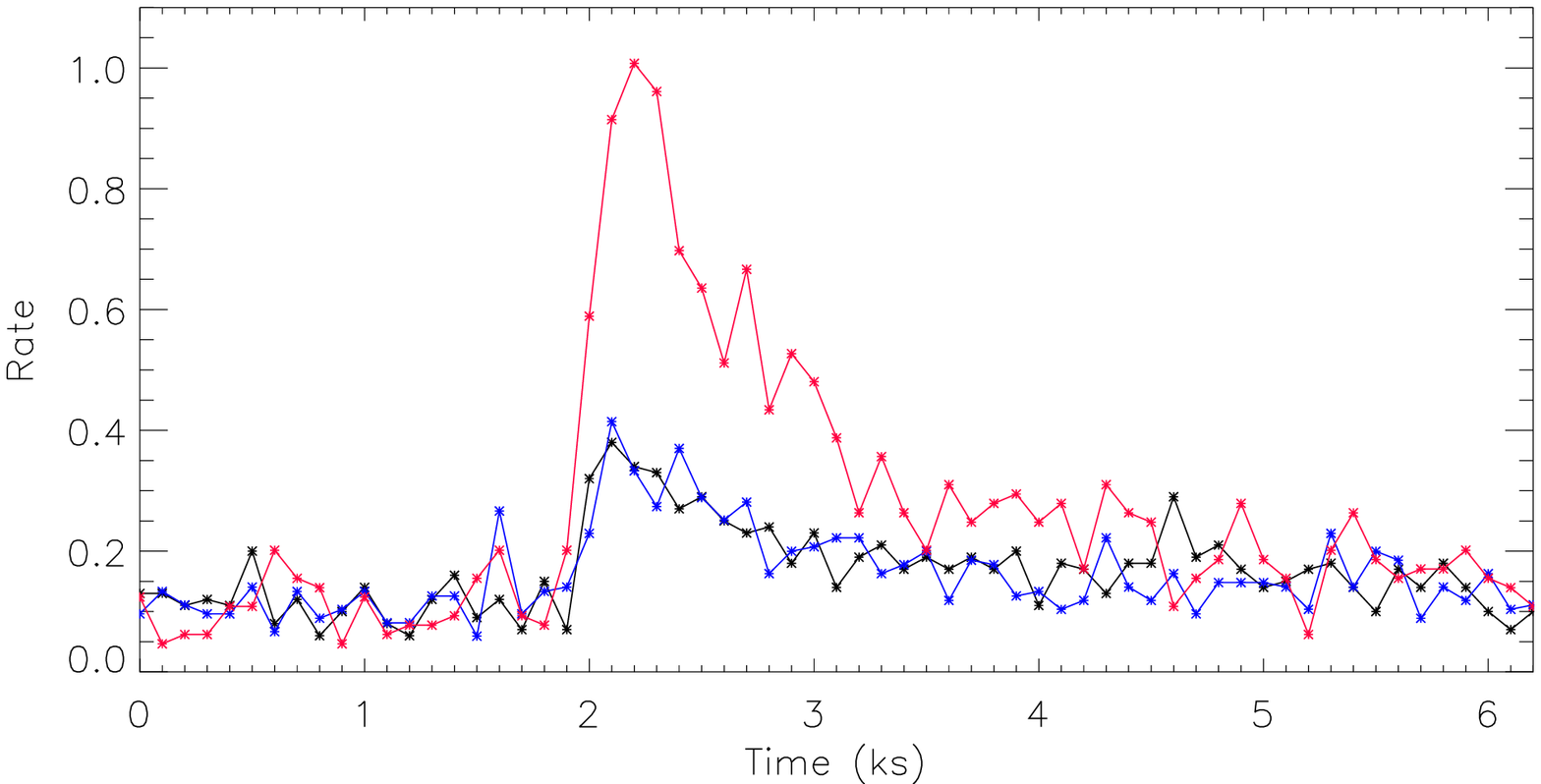}
\includegraphics[width=80mm]{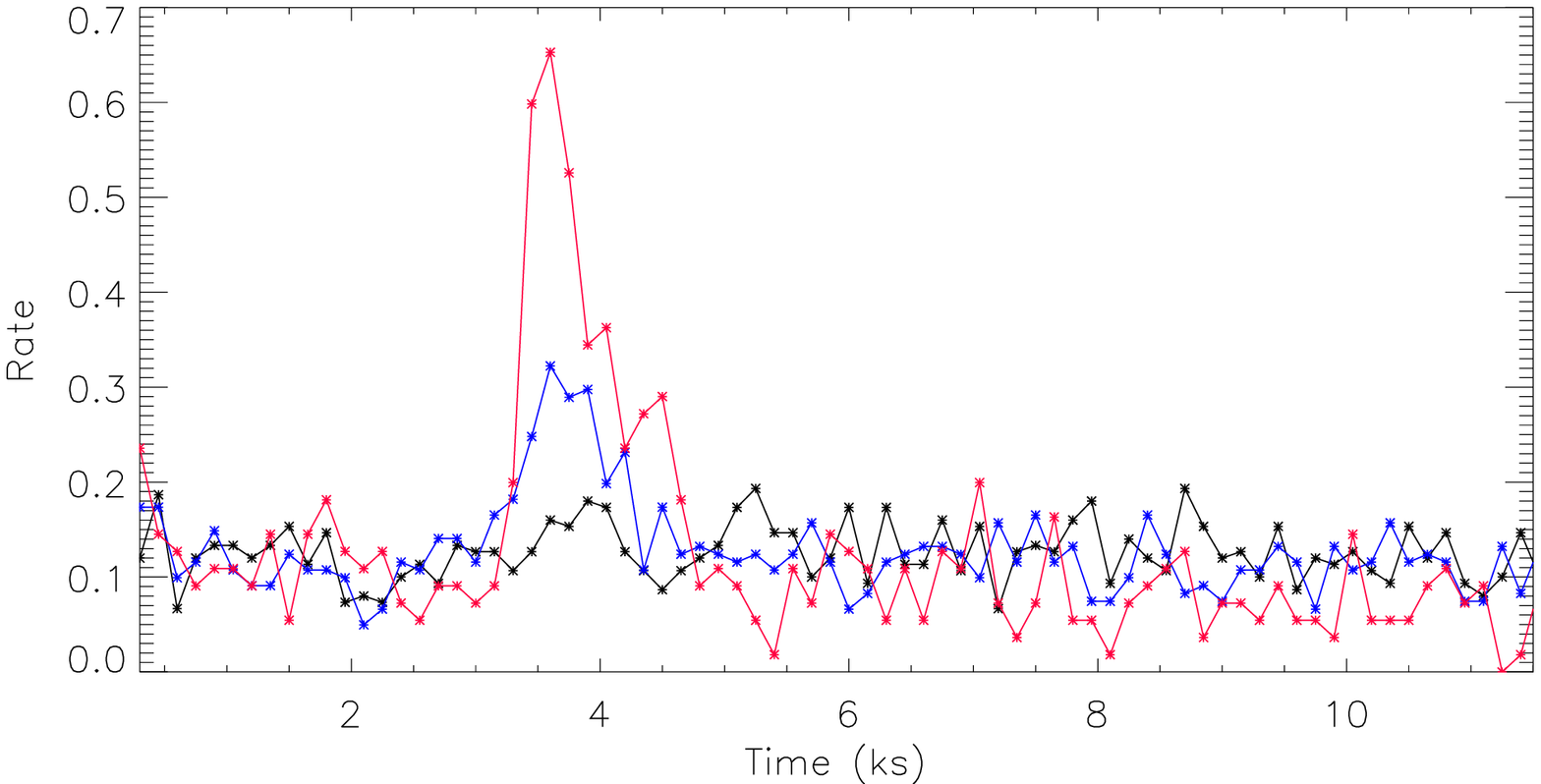}
\caption{\label{flcs}The respectively largest X-ray flares from 61\,Cyg~B (top), 61\,Cyg~A (middle) and $\alpha$\,Cen~B (bottom) observed during the XMM campaign;
PN data in the 0.2\,--\,0.5 (black), 0.5\,--\,0.8 (scaled, blue) and 0.8\,--\,5.0~keV (scaled, red) energy band.}
\end{figure}

The largest flare observed on each of the three mentioned stars is shown in Fig.\,\ref{flcs};
the strongest flare during the total campaign occurring on 61\,Cyg~B (11/2004), an intermediate one on 61\,Cyg~A (11/2003) and a small event on $\alpha$\,Cen~B (07/2008).
Due to the presence of hotter plasma we study the flares in the 0.2\,--\,5.0~keV band. The
shown X-ray light curves are subdivided into in three energy bands to trace the energetic behavior, here
the 0.5\,--\,0.8~keV and 0.8\,--\,5.0~keV band are scaled to match the respective pre-flare flux in the 0.2\,--\,0.5~keV band.
The temporal behavior of the $\alpha$\,Cen~B and 61\,Cyg~A flare are similar, both show short rise times of about 0.3~ks and two to three times longer decay times (e-fold) 
of 0.7 and 0.8~ks. In contrast, the 61\,Cyg~B flare with a rise time of 0.8~ks has a six times longer decay time of 5~ks; 
together with some flattening of the decay this indicates reheating or an erupting multiple loop structure.
Inspecting the flare light curves, the hardest spectral band always leads the flare rise and shows the by far largest relative flux increase, 
a typical behavior for solar and stellar flares in the chromospheric evaporation scenario. 
Note, that the shown $\alpha$\,Cen~B flare occurred around activity minimum and that the post-flare rate in 
the 0.8\,--\,5.0~keV is roughly 30\,\% lower compared to the pre-flare state;
this is in contrast to the medium and low energy band that stayed constant or even increase moderately, indicating that 
the active region producing flare is also attenuated by this event.

The $\alpha$\,Cen~B flare has a duration of roughly half hour, reaches at peak about $L_{\rm X} = 1.2 \times 10^{27}$~erg\,s$^{-1}$ and releases in total
$5 \times 10^{29}$~erg. 
A flare of similar energetics from $\alpha$\,Cen~B that was detected in 01/2004 around activity maximum is discussed in \cite{rob05};
likely these events are among the faintest stellar flares with a well observed rise and decay phase detected in X-rays.
Several small and intermediate events are also detected from 61\,Cyg~A and 61\,Cyg~B. 
The largest flare on
61\,Cyg~A occurred around its activity maximum in 2003 with an X-ray brightness increase by a factor of about five. It reaches a peak luminosity
of about $L_{\rm X} = 6 \times 10^{27}$~erg\,s$^{-1}$ and releases about $1 \times 10^{31}$~erg in soft X-rays within 3~ks.
The by far strongest flare event was observed on 61\,Cyg~B in 2004, here the X-ray brightness increases by a factor of about 20 compared to the quasi-quiescent state 
and flare plasma temperatures of 20~MK are present.
This flare reaches at maximum $L_{\rm X} = 1.4 \times 10^{28}$~erg\,s$^{-1}$ and releases in total $7 \times 10^{31}$~erg within 10~ks. 
As indicated by the MOS light curve shown in Fig.\,\ref{61lcs}, a second event occurred at the final stage of the flare decay, 
however only its initial rise phase is covered by the observation. 
As a comparison to large flares observed on the Sun, we transform the solar flux to the here used energy band.
A solar X class flare emits at peak above $2.8 \times 10^{26}$~erg\,s$^{-1}$ 
in the 1\,--\,8\,\AA\, (1.55\,--\,12.4~keV) GOES band, roughly corresponding to $1 \times 10^{27}$~erg\,s$^{-1}$ in the 0.2\,--\,5.0~keV band for a 15~MK plasma.

\section{Coronal activity cycles}
\label{cyc}
The investigation of cyclic activity uses the quasi-quiescent X-ray properties as determined for each star and observation.

\subsection{Long-term brightness variations}
The long-term evolution is derived from our {\it XMM-Newton} data and compared to measurements from other instruments; all properties refer to the 0.2\,--\,2.0~keV band.

\subsubsection{The 61 Cyg system}

\begin{figure}[t]
\centering
\includegraphics[width=90mm]{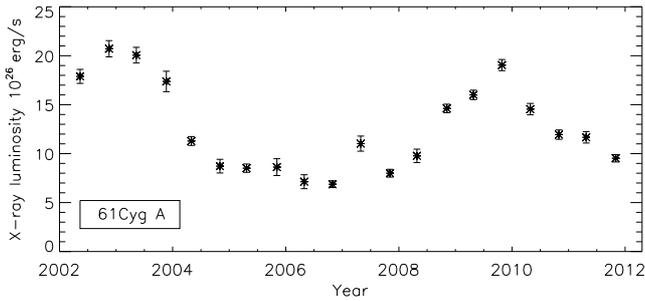}
\caption{\label{61cyga}The X-ray activity cycle of the K5 star 61\,Cyg A as observed with {\it XMM-Newton}.}
\end{figure}

The {\it XMM-Newton} observations of the 61\,Cyg system cover so far nearly 10 years and thus more than one complete activity cycle of 61\,Cyg~A 
with its chromospheric period of $7.3\pm 0.1$~yr.
As shown in Fig.\,\ref{61cyga}, the X-ray data exhibit quite smooth long-term variations and cover two distinct activity maxima. This is one of the rare cases where a clear
X-ray cycle on a star other than the Sun is covered for more than a period. Further, it refers to a mid K star
and illustrates the existence of stellar coronal activity cycles in a broader population in an unprecedented fashion.
The smoothness of the observed X-ray cycle also indicates, that effects induced by rotational modulation or variability on intermediate timescales of months
are only of minor importance.
The position of the first activity maximum in the new millennium was determined to be at mid/end 2002 from near simultaneous X-ray and chromospheric measurements \citep{hem06},
leading to the prediction of a new maximum occurring at the end of 2009. This is in good agreement with our X-ray measurements, 
indicating an identical X-ray cycle length within an accuracy of a few months. Expanding the time-line to the earliest chromospheric
observations, where a first minimum is observed around 1970.0 and combine it with our new X-ray maximum, we again recover the 7.3~yr period from \cite{bal95}
with an accuracy better than 0.1~yr over 40~years of observations. The combined chromospheric and coronal data of 61\,Cyg~A show that
the period of the activity cycle is quite stable over several decades and that no significant phase-shift exists between both activity indicators.

\begin{figure}[t]
\centering
\includegraphics[width=90mm]{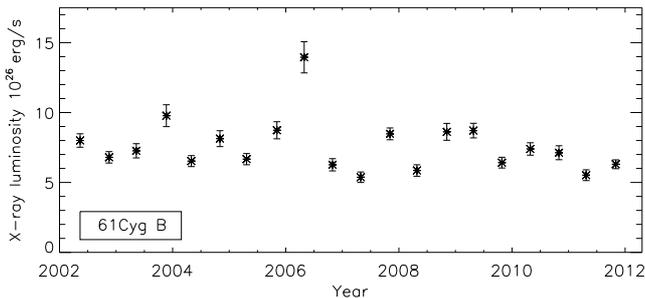}
\caption{\label{61cygb}The X-ray activity of the K7 star 61\,Cyg B observed with {\it XMM-Newton}, so far no cyclic long-term trend is present.}
\end{figure}

The second X-ray maximum is slightly more sharply peaked than the first one, but
overall the smooth, nearly sinusoidal X-ray brightness variations of 61\,Cyg~A are reminiscent of the solar activity cycle. A significant difference is
the amplitude ($L_{\rm Xmax}/ L_{\rm Xmin}$) between minimum and maximum activity,
which is only a factor of three on 61\,Cyg~A compared to a factor of roughly ten for the Sun at similar energies. 
The X-ray cycle of 61\,Cyg~A shows two X-ray maxima of comparable brightness and a rather broad or flat minimum with ongoing minor activity.
The X-ray luminosity varies between $L_{\rm Xmin} = 7 \times 10^{26}$~erg\,s$^{-1}$ and $L_{\rm Xmax} = 2.1 \times 10^{27}$~erg\,s$^{-1}$ over the cycle.
The first X-ray maximum is about 15\,\% higher than the second one, a rather minor difference given the intrinsic variability of the star.
Comparison with its optical luminosity results in a moderate activity level varying from $\log L_{\rm X}/L_{\rm bol} = -5.9\,... -5.4$,
thus 61\,Cyg~A is roughly an order of magnitude more active than the Sun. 
Very similar X-ray brightness levels were observed with {\it ROSAT} in the 1990s \citep{hem06} and
notably 61\,Cyg~A is with a spectral type of K5 also the latest stars with an 'excellent' activity cycle in the Mt. Wilson sample \citep{bal95}.

In contrast, the K7 star 61\,Cyg~B exhibits so far no clear cyclic X-ray brightness variations in the {\it XMM-Newton} observations (see Fig.\,\ref{61cygb}), 
despite that a chromospheric activity cycle
with a period of $11.7\pm 0.4$~yr has been derived and significant variability is present in the {\it ROSAT} data obtained over five years in the mid 1990s \citep{hem06}.
Extrapolating the chromospheric data, a new activity maximum should have occurred around 2004, but
remarkably the quasi-quiescent flux is rather constant and exhibits only a weak decline by 20\,\% as long-term trend
during our observations. 
Typical quasi-quiescent deviations from the mean brightness are less than 30\,\% and quite irregular, but several individual observations
are dominated by short-term behavior and weak to moderate flaring activity.
We find that 61\,Cyg~B resided at an intermediate state in a kind of activity plateau around
a mean X-ray luminosity of $L_{\rm X} = 8 \times 10^{26}$~erg\,s$^{-1}$, corresponding to an activity level of $\log L_{\rm X}/L_{\rm bol} = -5.6$.
Compared to the {\it ROSAT} data, where 61\,Cyg~B showed $L_{\rm X} = 4 \dots 10 \times 10^{26}$~erg\,s$^{-1}$, 
the observed X-ray luminosity is neither specifically low nor high.
This behavior is rather unexpected for a cyclic star, however
also its chromospheric cycle, although denoted as 'good' in \cite{bal95}, appears at times quite irregular and is less smooth than the one of 61\,Cyg~A.
Nevertheless, the X-ray data fits well to more recent \ion{Ca}{ii} observations of 61\,Cyg~B \citep{hall07}, which show a maximum around 2000 that is followed by a minor decline over about two years
and a rather flat activity since 2002. Altogether the data points to a slightly shorter 10\,--\,11~yr or variable activity cycle period.

\subsubsection{The $\alpha$\,Cen system}

\begin{figure}[t]
\centering
\includegraphics[width=92mm]{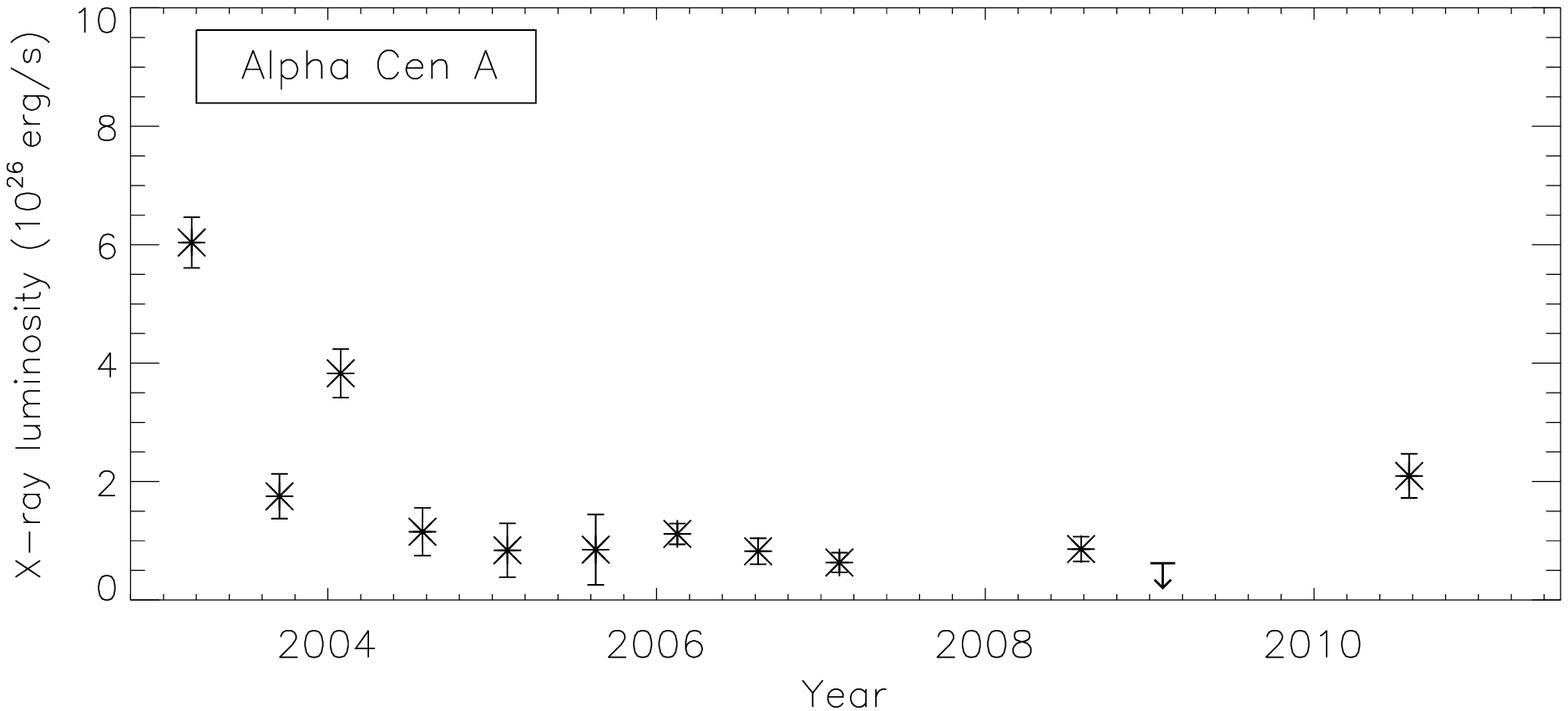}

\smallskip
\includegraphics[width=92mm]{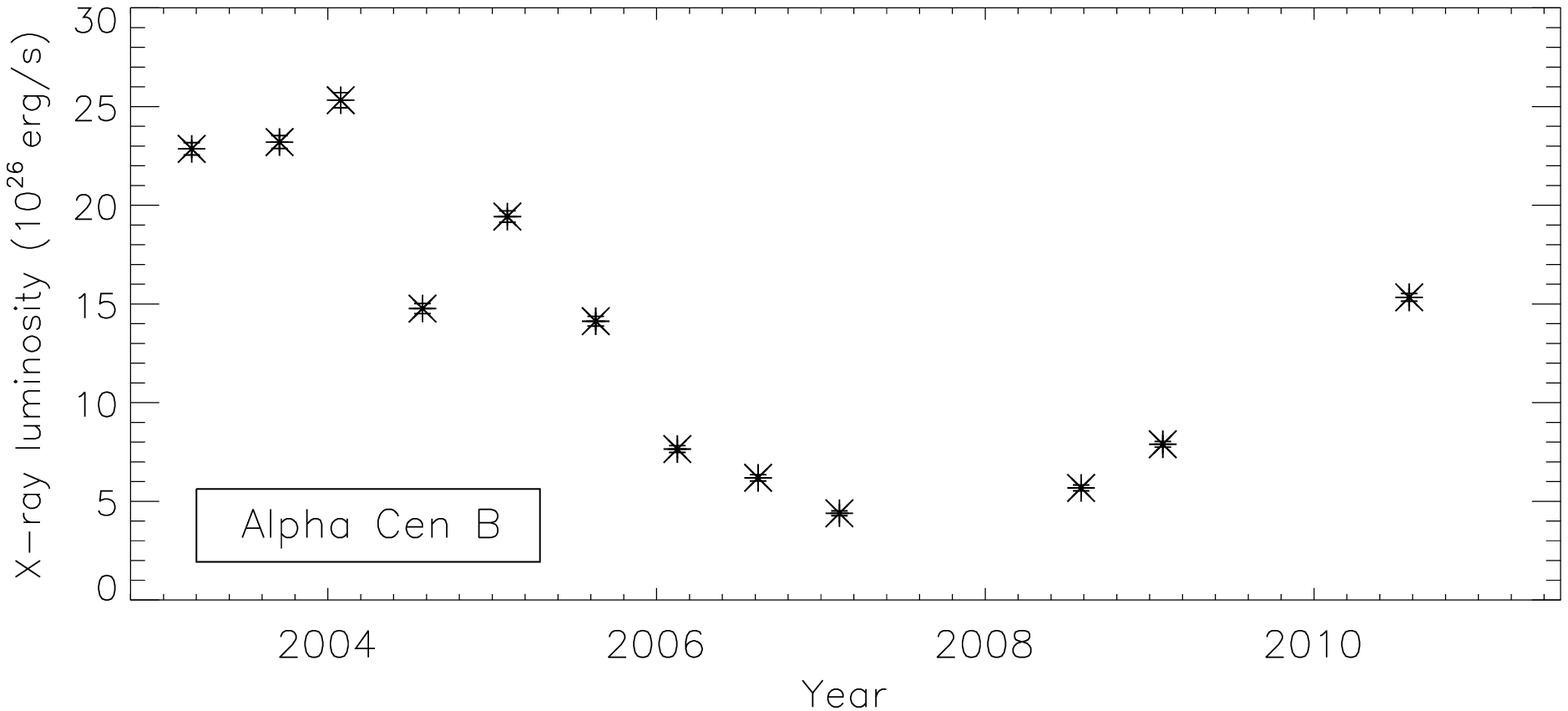}
\caption{\label{acenab}The X-ray activity of $\alpha$\,Cen~A (likely cyclic) and $\alpha$\,Cen~B (cyclic) as observed with XMM/MOS1 (A) and XMM/EPIC (B).}
\end{figure}

The earliest and least active star in our sample is the G2 star $\alpha$\,Cen~A. 
Attributing the observed activity trend over 7 years of X-ray monitoring
to the most likely scenario, i.e. an activity cycle, our observations indicate a cycle period of well above 10~yr with a rather broad minimum
as shown in Fig.\,\ref{acenab} (upper panel). This suggests that $\alpha$\,Cen~A exhibits an activity 
cycle that is similar to the solar one in respect of period and amplitude, but a longer observational time-line is clearly needed to confirm this hypothesis.
The relative X-ray brightness variations are the largest of the presented sample stars. 
The X-ray luminosity of $\alpha$\,Cen~A during the {\it XMM-Newton} observations ranges between $L_{\rm X min} \approx 7 \times 10^{25}$~erg\,s$^{-1}$ and
$L_{\rm X} = 6 \times 10^{26}$~erg\,s$^{-1}$; variations are roughly an order of magnitude, but the exact minimum is quite poorly constrained by these data. 
After reaching its faint state in 2005 $\alpha$\,Cen~A is 
barely detected at the $2-3 \sigma$ level with MOS1/PN and its luminosity stayed at a low but variable level but over the next five years.
We clearly detect $\alpha$\,Cen~A in several exposures from 2006 to 2010, but it
remained undetected in the 2009 MOS1 exposure where we obtain only an upper limit for its X-ray brightness.
Further, we have not covered an activity maximum, but considering the 30 year history of X-ray observations
the ratio between maximum and minimum X-ray brightness is about a factor of 15 and $\alpha$\,Cen~A has
a mean activity level of $\log L_{\rm X}/L_{\rm bol} \approx -7.0$.
If the observed decline is not an unprecedented event, we likely started observations after its last activity maximum and $\alpha$\,Cen~A has an coronal activity cycle
with at least an order of magnitude variability and a rough period estimation of $\sim 12 - 15$~yr.

The X-ray observing history of $\alpha$\,Cen~A helps to address the question of its long-term evolution and to cross-check our results.
As a check we also model the dispersed spectra of a second LETGS observation performed in 06/2007 analog to the MOS data and find $L_{\rm X} = 0.9 \times 10^{26}$~erg\,s$^{-1}$.
This is intermediate to our two nearest XMM values of $L_{\rm X} = 0.7/1.0 \times 10^{26}$~erg\,s$^{-1}$ (02/2007, 07/2008) and
indicates that the derived values reflect the flux of $\alpha$\,Cen~A in its low state quite well, i.e. $L_{\rm X min} \lesssim 1 \times 10^{26}$~erg\,s$^{-1}$.
As an additional test we sum up all measured LETGS lines
in the spectral range 6\,--\,60~\AA\, as given by \cite{ayr09}, 
treat all upper limits as detections and account for the missing flux by using APEC models (see Table~\ref{apecc}). 
The plasma model uses the FIP pattern abundances as described in Sect.\,\ref{obsana} and results are checked with a solar abundance model.
For appropriate plasma temperatures, here around 1~MK, we obtain with this approach 
$L_{\rm X} \approx 1.2- 1.4 \times 10^{26}$~erg\,s$^{-1}$.
To trace its evolution backwards in time, we
repeat the spectral modeling  and line flux conversion for the 12/1999 LETGS observation as above 
and find an X-ray luminosity of $L_{\rm X} \approx 3.5 - 4.5 \times 10^{26}$~erg\,s$^{-1}$.
These results also fit to the $L_{\rm X} \approx 2 \times 10^{26}$~erg\,s$^{-1}$ derived by \cite{dew10} for the {\it ROSAT} 1996 data,
supporting the idea of an activity cycle, where
the last $\alpha$\,Cen~A maximum has occurred early in the new millennium before the start of our {\it XMM-Newton} observations.

The {\it XMM-Newton} observations of the K1 star $\alpha$\,Cen~B do not yet cover a full cycle, but the long-term trend
in X-ray luminosity exhibits a likely maximum at end 2003/early 2004,
a clear minimum around end 2007/early 2008 and a successive rise until 2010 (see Fig.\,\ref{acenab}, bottom panel).
The observed trend in X-rays fully supports an activity cycle with a period of about 8\,--\,9~yr as derived from a combined analysis of multiple activity indicators \citep{dew10}.
Our X-ray data indicates X-ray brightness variations over the cycle of roughly a factor six to eight in the 0.2\,--\,2.0~keV band. X-ray luminosities
range between $L_{\rm X min} \approx 4 \times 10^{26}$~erg\,s$^{-1}$ and $L_{\rm X max} = 2.5 \times 10^{27}$~erg\,s$^{-1}$,
corresponding to a mean activity level of $\log L_{\rm X}/L_{\rm bol} = -6.1$. Calculating as above the LETGS flux value for the 06/2007 observation, we find 
$L_{\rm X} \approx 3.0 - 4.0 \times 10^{26}$~erg\,s$^{-1}$ from spectral modeling and emission line flux conversion, 
in good agreement with expectations from the 02/2007 {\it XMM-Newton} near minimum value of $L_{\rm X} = 4 \times 10^{26}$~erg\,s$^{-1}$.
Our data indicate an activity maximum at the beginning of the {\it XMM-Newton} monitoring, 
thus one expects a previous activity minimum to have occurred around the end of the 1990s.
This is supported by the LETGS data from 1999, where we derive an X-ray luminosity of
$L_{\rm X} \approx  4.0 - 5.0 \times 10^{26}$~erg\,s$^{-1}$ (note that the $L_{\rm X}$ in \cite{raa03} for these data are overestimated, priv. comm.),
similar to our minimum $L_{\rm X}$ from the 2007 data.
Extrapolating further backwards in time, it is not surprising that {\it Einstein} \citep[1979, $L_{\rm X} = 2.8 \times 10^{27}$~erg\,s$^{-1}$,][]{gol82}
and {\it ROSAT} \citep[1995/96,  $L_{\rm X} \approx 2 \times 10^{27}$~erg\,s$^{-1}$,][]{dew10} measured an X-ray brightness
similar to the {\it XMM-Newton}  (2004, $L_{\rm X} = 2.5 \times 10^{27}$~erg\,s$^{-1}$) values in the beginning of our campaign; by chance all three satellites observed $\alpha$\,Cen~B around its activity maximum.

This paper focusses on the {\it XMM-Newton} data and we find an overall agreement between 0.2\,--\,2.0~keV luminosities derived from the MOS and PN detectors, 
independent if they are obtained from spectral modeling or by conversion of count rates.
We also find consistency to LETGS spectra and for the combined $\alpha$\,Cen system to the RGS, when modeling these data, converting emission line fluxes
or comparing flux ratios of strong lines.
We note that discrepancies exist between literature values for the {\it ROSAT} HRI data and
that fluxes obtained from contemporaneous {\it Chandra} HRC-I observations \citep{ayr09} are above our results;
a similar inconsistency exists between {\it Chandra} ACIS-S and HRC-I observations of 51~Peg \citep{pop09}.

\subsection{Spectra and spectral changes}

In the following we investigate the spectra of our target stars and study changes of their coronae over the activity cycles.

\subsubsection{61 Cygni A/B}

\begin{figure}[t]
\includegraphics[height=85mm,angle=-90]{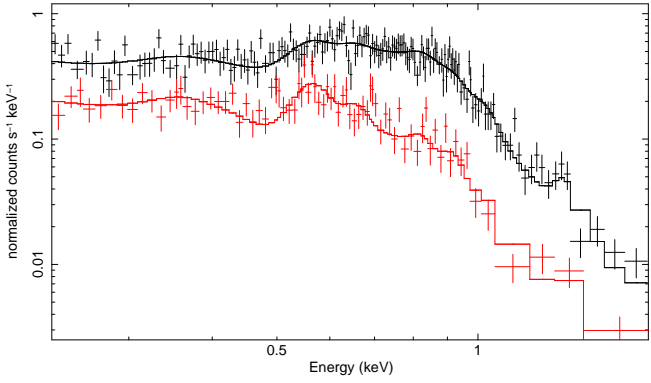}
\caption{\label{specs}PN spectra of 61 Cyg A around maximum (black, 10/2009) and minimum (red, 10/2006) coronal activity with spectral models.}
\end{figure}

To characterize the cycle properties of 61\,Cyg~A, we study observations around maximum and minimum X-ray luminosity
where we simultaneously model three observations for each state.
We determine the respective coronal temperatures and emission measure distributions and as above restrict the analysis to the quasi-quiescent state.
To account for minor X-ray brightness differences of about 10\,\% between individual observation,
the normalization is taken as a free parameter and the result is averaged for each activity state.
Here we use data of the more sensitive PN detector, but qualitatively very similar results are obtained for the MOS data.
As an example we show two PN spectra representative of the two activity states with their respective model in Fig.\,\ref{specs};
the obtained modeling results for 61\,Cyg~A are summarized in Table\,\ref{specres}.

\begin{table}[t]
\centering
\caption{\label{specres}Corona of 61\,Cyg A at maximum and minimum activity.}
\begin{tabular}{lrrrr}\hline\hline\\[-3mm]
Par.&  max. & min. & ratio \\\hline\\[-3mm]
$kT_{1}$ & 0.10 $\pm 0.01$ &  0.09 $\pm 0.01$ & &keV\\
EM$_{1}$  & 9.54 $\pm 1.40$& 5.79 $\pm 1.08$ & 1.6 &  $10^{49}$\,cm$^{-3}$\\
$kT_{2}$  & 0.33 $\pm 0.02$& 0.30 $\pm 0.02$  & &keV\\
EM$_{2}$  & 4.22 $\pm 0.43$& 1.53 $\pm 0.14$ & 2.8 &$10^{49}$\,cm$^{-3}$\\
$kT_{3}$  & 0.69 $\pm 0.06$ & 0.68 $\pm 0.29$ & &keV\\
EM$_{3}$  & 1.12 $\pm 0.38$ & 0.11 $\pm 0.11$& 10. &$10^{49}$\,cm$^{-3}$\\\hline\\[-3mm]
$\chi^2_{\rm red}${\tiny(d.o.f.)} & 1.05 (396)& 1.08 (166) &&\\\hline\\[-3mm]
av. $L_{\rm X}$ & 1.9 &0.7 & &$10^{27}$\,erg s$^{-1}$\\\hline
\end{tabular}
\end{table}

The multi-temperature models indicate an average plasma temperature of about 1.5\,--\,2.0~MK around the activity minimum and
of roughly 2.5~MK around maximum; for example the models shown in Fig.\,\ref{specs} correspond to average coronal temperatures of 1.6~MK and 2.4~MK.
Comparing the individual plasma components at maximum and minimum activity, we find that their temperatures are virtually identical, but the emission measure
changes significantly and further in a temperature dependent way. In comparison to the Sun, the three components can be associated
with main solar coronal structures; i.e. 1\,--\,2~MK corresponding to the quiescent Sun (network, coronal holes), 2\,--\,5~MK to active regions 
and hotter plasma to flaring regions \citep{orl01}.
The hot plasma component at 8~MK is at activity maximum about ten times, the intermediate component at 4~MK is roughly 2-3 times 
and the cool component at 1~MK is only about 50\,\% stronger than at minimum.
However, 61\,Cyg~A is only moderately active and even around maximum the 1~MK component contributes about 60\,\% to the total EM, 
whereas around minimum it is up to 80\,\%. 
While the 8~MK plasma component is only marginally detected around activity minimum, significant amounts of 4\,--\,5~MK plasma are always present.
Thus a state similar to the 'quiet Sun', i.e.
a 1\,--\,2~MK corona without any significant contribution from hotter plasma is never present on 61\,Cyg~A.

Despite the facts that the cycle period is significantly shorter, that the X-ray luminosity varies only by a factor of three and that 61\,Cyg~A is generally
more active than the Sun, the overall coronal trends are identical for both activity cycles. Given the smoothness and regularity of
the activity cycle on 61\,Cyg~A, these findings indicate that the underlying mechanisms generating cyclic activity are also quite similar.

Like the X-ray brightness of 61\,Cyg~B, also its quasi-quiescent spectra are quite similar for all observations. 
The corona of 61\,Cyg~B is well described by a plasma distribution with similar temperature components as found for the primary. 
The corona of 61\,Cyg~B is with an average temperature of about 2.2~MK moderately hotter than those of 61\,Cyg~A
when comparing phases of similar X-ray brightness, but the coronal temperature distributions become quite similar when comparing identical activity levels.

\subsubsection{$\alpha$ Centauri A/B}

\begin{table}[t]
\centering
\caption{\label{specres2}Coronal EMD of $\alpha$\,Cen B over the activity cycle.}
\begin{tabular}{lrrrrr}\hline\hline\\[-3mm]
Par.& 09/2003  &  02/2007 & & 2010 \\
&  (max.) &   (min.) & ratio& \\\hline\\[-3mm]
$kT_{1}$ & 0.12 $\pm 0.01$ & 0.11 $\pm 0.01$& &0.12 &keV\\
EM$_{1}$ & 14.8 $\pm 1.7$ & 3.9 $\pm 0.5$&3.8 &11.1 & $10^{49}$\,cm$^{-3}$\\
$kT_{2}$ & 0.28 $\pm 0.02$& 0.28 $\pm 0.03$& &0.26 &keV\\
EM$_{2}$ & 6.9 $\pm 1.0$& 0.8 $\pm 0.3$ &8.6 & 3.9  &$10^{49}$\,cm$^{-3}$\\
$kT_{3}$ & 0.68 $\pm 0.12$  & -- & &0.55 &keV\\
EM$_{3}$ & 0.4 $\pm 0.2$ & -- & $>$15 &0.3 & $10^{49}$\,cm$^{-3}$\\\hline\\[-3mm]
$\Sigma$ EM & 22.1  & 4.7 & &15.3 & $10^{49}$\,cm$^{-3}$\\
av. $T_{\rm X}$ & 2.1 & 1.6 && 1.9 & MK\\
$L_{\rm X}$ &2.2 & 0.4 &&1.4 &$10^{27}$\,erg s$^{-1}$\\\hline
\end{tabular}
\end{table}

Spectra of the $\alpha$\,Cen A/B system show that the coronae are strongly dominated by plasma at relatively cool temperatures at all phases. 
At maximum X-ray brightness in 2003 the spectra are well described by a 3-T model with temperatures of about 1.5, 3.5 and 8.0~MK.
The EMD is dominated by 1\,--\,4~MK plasma,
resulting in an average temperature of 2\,--\,2.5~MK. Separate spectral modeling shows that $\alpha$\,Cen~A exhibits a moderately larger EM1/EM2 ratio for
the two cooler components, whereas the 8~MK component is significantly present only on $\alpha$\,Cen~B.
Thus the X-ray brighter K1 star $\alpha$\,Cen~B exhibits a hotter corona than the G2 star $\alpha$\,Cen~A, 
during all of our observations.
The X-ray luminosity of $\alpha$\,Cen~A showed a quite rapid drop in within two years and in later observations
photons are detected almost exclusively at very soft X-ray energies ($< 0.5$~keV),
indicating dominant temperatures of $\lesssim 1$~MK and a virtually complete vanishing of hotter coronal structures.

The secondary $\alpha$\,Cen~B strongly dominates the system especially in all later {\it XMM-Newton} observations
and we model its coronal properties for one representative observation around maximum, minimum and intermediate activity phase.
Typical results, here obtained from the FIP model for the MOS data, are given in Table\,\ref{specres2}. 
The modeled temperatures and emission measures for individual exposures show changes over the activity cycle that resemble those of 61\,Cyg~A.
The cool 1.5~MK plasma component dominates the emission at all times; even around activity maximum it contributes above 60\,\% of the total emission measure
and is roughly twice as strong as the 3.5~MK component. In addition,
around maximum a weak 8~MK component that contributes by a few percent to the total emission measure is detected in the quasi-quiescent phase. Towards activity minimum
the 8~MK component vanishes and a strong decline in emission measure occurs in the other two components.
The decline of the 3.5~MK component is with a factor of nearly ten more than twice as strong as those of the 1~MK component, 
leading to a decline of the average coronal temperatures to around 1.5~MK at minimum. 
During the re-brightening of $\alpha$\,Cen~B in 2010 a hotter plasma component appears, 
but it is still slightly cooler and weaker than during activity maximum.

\subsection{X-ray cycles of the sample stars vs. the Sun}
\label{sun}

We discuss our findings on coronal activity cycles in a broader context and compare them to stellar and solar activity cycles as seen in X-rays and other activity indicators.

The basic stellar and X-ray activity cycle properties of the presented stars are summarized in Table\,\ref{xcy}.
Stellar data are taken from references given in Sect.\,\ref{tar} and X-ray properties are from this work;
literature values are given in 'italic' when the activity range is not fully covered. 
In addition to the here studied stars, we use as comparison two further G2 stars, HD~81809 and the Sun.
The moderately active star HD~81809 is also part of our observing program \citep{fav08}. It has a chromospheric cycle of 8.2~yr and an X-ray luminosity range
of $L_{\rm X} \approx 1.4 - 6.4 \times 10^{28}$~erg\,s$^{-1}$ in quasi-quiescence, corresponding to an average $\log L_{\rm X}/L_{\rm bol} = -5.6$.
The Sun exhibits an average cycle period of 11~yr and has an X-ray luminosity range
of $L_{\rm X} \approx 0.5 - 6 \times 10^{27}$~erg\,s$^{-1}$, corresponding to an average $\log L_{\rm X}/L_{\rm bol} = -6.2$ \citep[see e.g.][]{jud03}.

\begin{table}[t]
\centering
\caption{\label{xcy}Stellar and X-ray cycle properties (0.2\,--\,2.0~keV) of the analyzed stars; brackets denote literature values, $\log L_{\rm X}$ in [erg\,s$^{-1}$].}
\begin{tabular}{lrrrr}\hline\hline\\[-3mm]
Par.&  $\alpha$\,Cen A & $\alpha$\,Cen B & 61 Cyg A & 61 Cyg B\\\hline\\[-3mm]
sp. type& G2V & K1V & K5V & K7V \\
$P_{\rm rot}$ (d) & 29 & 37 & 35 & 38\\
$P_{\rm act}$ (yr) & $\sim$15?& 8.8 & 7.3 & 11.7\\
$\log L_{\rm X max}$& 26.8 ({\it 27.1}) & 27.4 & 27.3 & 27.1\\
$\log L_{\rm X min}$& $\lesssim$26 & 26.6 & 26.8 & 26.7 ({\it 26.6})\\
$\log L_{\rm X}/L_{\rm bol}$ & $-$7.0 & $-$6.2& $-$5.6 &$-$5.5\\
X-ray cycle & Y? & Y & Y &Y? \\
Ampl. & $\gtrsim$10 & 6 & 3 & 2\,--\,3?\\
\hline
\end{tabular}
\end{table}

Sorting the three G2 stars by increasing activity, i.e. $\alpha$\,Cen~A, Sun and HD~81809 we find as a general
trend for their coronal activity cycles that
the less active stars have activity cycles with longer periods and larger relative brightness variation at soft X-ray energies above 0.2~keV.
Comparing stars with a similar activity level, i.e. $\alpha$\,Cen~B vs. Sun and 61\,Cyg~A vs. HD~81809, we find that
the stars of earlier spectral type have longer periods and larger brightness variations at soft X-ray energies.
These trends are based on a few stars only, but might be representative for coronal activity cycles in general.

A common feature of the studied stars are the variations of their coronal properties over the activity cycle.
The observed changes in the emission measure distribution of 61\,Cyg~A and  $\alpha$\,Cen~B 
are similar to the ones that are present in the solar corona and on HD~81809, despite different underlying stars and activity levels.
In all studied coronae the variations in emission measure basically determine the amplitude of the cycle.
Further the emission measure reduction is significantly stronger in the respective hotter plasma components and
as a consequence, an X-ray luminosity decrease is always accompanied by a decrease of the average coronal temperature.
In addition,
the absolute temperature scale of this trend and the individual emission measure distributions depend on the activity level of the respective star.
For example, $\alpha$\,Cen~B and 61\,Cyg~A show similar trends, but they are shifted towards cooler temperatures in lesser active star.
As a consequence cycle amplitudes are strongly dependent on the X-ray energy band and
specifically an X-ray cycle has more pronounced brightness contrast between activity maximum and minimum in harder X-ray bands, that - if sufficiently broad - 
typically trace increasingly hotter plasma.
Beside that the cyclic stars are always weakly or at best moderately active, they do not differ from other magnetically active stars, 
i.e. X-ray brighter stars have higher coronal temperatures and exhibit stronger and more frequent X-ray variability and flaring.
The activity scaling and similarity between cycle properties indicate, that 
the changes of coronal properties over an activity cycle base on similar physical scenarios and follow a universal pattern in all stars.

It is instructive to compare the activity minimum state of our sample stars to other inactive stars as observed with ROSAT (0.1\,--\,2.4~keV)
and the Sun, where \cite{schmitt97} 
find that the minimum surface flux is well described by the flux from 'quiet Sun' regions as derived from soft X-ray observations of solar coronal holes,
there given as $\log F_{\rm X} = 4.1 \pm 0.2$~erg\,cm$^{-2}$\,s$^{-1}$.
The full Sun has at activity minimum, adapting $\log L_{\rm X} =26.8$~erg\,s$^{-1}$ \citep{jud03} an average surface flux of
$\log F_{\rm X} = 4.0$, i.e. slightly above the minimum value of $\log F_{\rm X} = 3.9$. 
We convert our X-ray measurements to ROSAT band fluxes by using the parameters given in Table~\ref{apecc} and obtain
minimum surface fluxes of $\log F_{\rm X} = 4.4$~erg\,cm$^{-2}$\,s$^{-1}$ for 61\,Cyg~A and B, $\log F_{\rm X} = 3.3$ for $\alpha$\,Cen~A and $\log F_{\rm X} = 4.0$ for $\alpha$\,Cen~B.
61\,Cyg~A has at activity minimum a higher surface flux than the minimum Sun. This indicates that some relevant and more active regions are also present
during the minimum of 61\,Cyg~A, in agreement with our findings from spectral modeling.  Comparable results are obtained for the observed phases of 61~Cyg~B.
The minimum surface flux of $\alpha$\,Cen~B is quite similar to the solar one and with $T_{\rm X} = 1.5$~MK it also exhibits similar temperatures, 
thus it is likely a fair 'coronal minimum twin'.
For $\alpha$\,Cen~A we have adopted above a low coronal temperature of 1~MK, i.e. assume twice as much flux in the 0.1\,--\,2.4~keV band, but still
one would need a five times higher X-ray flux to obtain a value of $\log F_{\rm X} = 3.9$~erg\,cm$^{-2}$\,s$^{-1}$. Significantly lower coronal temperatures
seem to be ruled out, inspecting the EMD of $\alpha$\,Cen~A  in \cite{ayr09}.
While there is some uncertainty in the minimum $L_{\rm X}$, the required increase is quite high and 
by using a 50\,\% higher surface flux we still derive only $\log F_{\rm X} = 3.4$.
Thus if accepting the above solar values, $\alpha$\,Cen~A at minimum is less active than the minimum Sun
and its average surface flux is even below the flux from solar coronal holes.
Alternatively, the solar minimum flux derived in most studies is not correct and e.g.
\cite{orl01} find $\log F_{\rm X} = 3.3 \dots 3.5 $~erg\,cm$^{-2}$\,s$^{-1}$ in the {\it ROSAT} band
for low activity data from the mid 1990s, very similar to our $\alpha$\,Cen~A results. We caution that solar minimum flux estimates and cycle amplitudes
from different modeling approaches easily vary by an order of magnitude.
These discrepancies re-emphasize the need of proper disk integrated solar X-ray fluxes over its activity cycle, 
but a 0.1\,--\,2.4~keV minimum surface flux limit of $\log F_{\rm X} \approx 4.0$~erg\,cm$^{-2}$\,s$^{-1}$ 
for magnetically active late-type stars is apparently beaten down by the least active stars.

Comparing the cycles properties observed by coronal indicators (X-rays) with those of chromospheric indicators (\ion{Mg}{ii}, \ion{Ca}{ii} HK)
and again specifically 61\,Cyg~A to the Sun, we find a mean S-index of $\langle S \rangle = 0.63 - 0.66$ for 61\,Cyg~A and of $\langle S \rangle = 0.17- 0.18$ for the Sun.
However, the S-index depends on the reference continuum and thus on the underlying star.
When subtracting the photospheric component one obtains $\log R\,^{\prime}_{\rm HK}= -4.94$ for the Sun and $\log R\,^{\prime}_{\rm HK}= -4.76$ for 61\,Cyg~A \citep{hall07}, 
thus the coronally more active star 61\,Cyg~A as reflected by a larger $\log L_{\rm X}/L_{\rm bol}$ value is also chromospherically more active.
When comparing the chromospheric excess fluxes, i.e.  corrected for photospheric and 'basal' contribution given as $\langle \Delta F_{\rm Ca} \rangle$ in \cite{hall07}, 
one finds comparable values for both stars, very similar as their absolute X-ray luminosities. 
The deviation of the seasonal mean in \ion{Ca}{ii} flux is for the Sun (0.144) again slightly larger than those of 61\,Cyg~A (0.124), 
but cyclic variations are clearly less pronounced than in X-rays. 
The $\alpha$\,Cen binary stars are part of the CTIO \ion{Ca}{ii} observing program of southern solar-like stars 
and \cite{hen96} give for two periods of data taking in 1992/1993 mean values of
$\langle S \rangle = 0.16$ and $\log R\,^{\prime}_{\rm HK}= -5.00$ for $\alpha$\,Cen~A and $\langle S \rangle = 0.21$ and $\log R\,^{\prime}_{\rm HK}= -4.92$ for $\alpha$\,Cen~B.
As in X-rays, $\alpha$\,Cen~A is also in \ion{Ca}{ii} less active than the Sun, whereas $\alpha$\,Cen~B is slightly more active.
\cite{buc08} derive \ion{Ca}{ii} S-indices from ground based data obtained in 2002\,--\,2004 at the CASLEO observatory and
derive values of $\langle S \rangle \approx 0.15$ for $\alpha$\,Cen~A and $\langle S \rangle \approx 0.22$ for $\alpha$\,Cen~B;
at these times $\alpha$\,Cen~A was declining towards it low activity state, while $\alpha$\,Cen~B was around its X-ray maximum.

Overall, coronal activity cycles are, like their chromospheric analogs, a quite common feature in weakly active G and K dwarfs.
The similarity of the coronal changes observed in stars with spectral types ranging from G2 to K5 and different activity levels 
suggests that an universal underlying dynamo mechanisms is causing the activity cycles in all stars, where spectral type and
and rotation determine the respective activity level and cycle properties.

\section{Summary}
\label{summ}

Significant long-term, most likely cyclic X-ray variability is detected in all four studied
early G to mid K dwarfs. Our sample stars have weak to moderate activity levels
($\log L_{\rm X}/L_{\rm bol} \sim -5.5\dots -7$), on which
the absolute coronal temperature scale depends. However, all coronal cycles show solar-like variations, 
i.e. they are more dominant in respective hotter plasma components.
X-ray luminosities vary by up to an order of magnitude at soft X-ray energies above 0.2~keV, thus monitoring is essential
to determine the true activity range for these stars.

The K5 star 61\,Cyg~A shows a smooth and persistent coronal activity cycle. It is
the latest and most active star with an detected regular X-ray cycle in our sample.
The X-ray cycle amplitude is with a factor of about three significantly smaller than for the Sun, but the changes of coronal 
properties over its 7~yr cycle are very similar.

The K7 star 61\,Cyg~B shows a weakly declining X-ray trend that 
is likely part of an more irregular 11~yr cycle. 
It is the most active star in our sample and
exhibits the most frequent short-term variability and strongest flaring activity.

The G2 star $\alpha$\,Cen~A is the least active star in our sample and shows with a factor of about ten the strongest X-ray variability.
No complete cycle has been covered by our monitoring, but
the long-term X-ray observations indicate cyclic activity with an order of magnitude variations at soft X-ray energies and a likely period around 12\,--\,15~yr, 
i.e. a slightly longer and stronger cycle than observed on the Sun.

The K1 dwarf $\alpha$\,Cen~B exhibits quite smooth X-ray variability of roughly a factor six to eight between maximum and minimum activity and our observations support a activity cycle with a
period of 8\,--\,9 years. Again, the coronal changes over the cycle resemble the solar behavior.

\begin{acknowledgements}
This work is based on observations obtained with XMM-Newton, an ESA science
mission with instruments and contributions directly funded by ESA Member
States and the USA (NASA).\\
J.R. acknowledges support from DLR under 50QR0803.
\end{acknowledgements}

\bibliographystyle{aa}
\bibliography{19046}


\begin{appendix}

\section{Observation log and applied conversion factors}

\begin{table*}[t]
\caption{\label{allobs}Observation log of data used in this publication.}
\begin{center}
\begin{tabular}{lccc}\hline\hline\\[-3mm]
Obs.ID & Instr. Mode &  Date  & Obs. Time MOS/PN (ks)\\\hline\\[-3mm]
 \multicolumn{4}{c}{61 Cygni}\\\hline\\[-3mm]
0041740101 & LW/thick & 2002-05-14 & 0.6/5.0\\
0041740301 & LW/thick & 2002-11-17 & 9.3/7.6 \\
0041740901 & LW/thick & 2003-05-12 & 7.7/6.2\\
0041741001 & LW/thick & 2003-11-22 & 8.2/6.7 \\
0041741101 & LW/thick & 2004-05-01 & 14.2/12.8\\
0043630101 & LW/thick & 2004-11-02 & 16.0/14.0 \\
0043630401 & LW/thick & 2005-04-23 & 9.7/7.7 \\
0202610401 & FF/thick & 2005-11-05 & 5.2/3.5 \\
0202610501 & FF/thick & 2006-04-29 & 5.2/3.5 \\
0401880801 & FF/thick & 2006-10-29 & 15.0/13.0\\
0401880901 & FF/thick & 2007-04-30 & 12.0/10.0\\
0501930601 & FF/thick & 2007-11-03 & 14.7/13.1\\
0501930701 & FF/thick & 2008-04-26 & 11.7/10.0\\
0550061301 & FF/thick & 2008-11-06 & 16.5/14.8\\
0550060801 & FF/thick & 2009-04-27 & 11.7/10.0\\
0600820801 & FF/thick & 2009-10-27 & 14.7/13.0\\
0600820901 & FF/thick & 2010-04-28 & 13.7/12.0\\
0654550601 & FF/thick & 2010-10-31 & 12.7/11.0\\
0654550701 & FF/thick & 2011-04-24 & 10.7/9.0\\
0670320201 & FF/thick & 2011-11-02 & 12.7/11.0\\\hline\\[-3mm]
 \multicolumn{4}{c}{$\alpha$ Centauri}\\\hline\\[-3mm]
0045340901 & SW/thick & 2003-03-04 & 6.8/6.7\\
0045341001 & SW/thick & 2003-09-15 & 7.6/7.4\\
0045341101 & SW/thick & 2004-01-29 & 5.2/5.0\\
0045340401 & SW/thick & 2004-07-29 & 7.7/7.5\\
0143630501 & LW/thick & 2005-02-01 & 8.8/4.6\\
0143630201 & LW/thick & 2005-08-17 & 7.0/5.0\\
0202611201 & SW/thick & 2006-02-15 & 9.1/8.9\\
0202611301 & SW/thick & 2006-08-13 & 8.7/8.5\\
0202611401 & SW/thick & 2007-02-09 & 12.5/12.3\\
0550060901 & FF/thick & 2008-07-30 & 14.0/12.3\\
0550061001 & FF/thick & 2009-01-27 & 13.7/12.0\\
0654550801 & FF/thick & 2010-07-29 & 15.7/14.0\\\hline
\end{tabular}

\end{center}
\end{table*}

\begin{table*}[t]
\caption{\label{apecc}Conversion factors from APEC models in XSPEC/PIMMS.}
\begin{center}
\begin{tabular}{lcccccccc}\hline\hline\\[-3mm]
 \multicolumn{9}{c}{Ratio of energy band flux to 0.2\,--\,2.0 keV flux}\\\hline\\[-3mm]
T (MK) & 0.8 & 1.0 & 1.2 & 1.4 & 1.6 & 1.8 & 2.0 &2.2 \\\hline\\[-3mm]
0.15 - 4.0 keV& 2.05 & 1.52 &1.26 & 1.14 & 1.09 &1.09 &1.09 & 1.10\\
0.10 - 2.4 keV & 3.25 & 1.88& 1.43 & 1.24 & 1.17 &1.16& 1.16 & 1.17\\\hline\hline\\[-2mm]
\multicolumn{9}{c}{MOS1 count rate to 0.2\,--\,2.0 keV flux (cts\,s$^{-1} =>$ erg\,cm$^{-2}$\,s$^{-1}$)}\\\hline\\[-3mm]
T (MK) & 0.8 & 1.0 & 1.2 & 1.4 & 1.6 & 1.8 & 2.0 & 2.2\\\hline\\[-3mm]
Conv. factor & 2.15$\cdot 10^{-11}$ &  1.71$\cdot 10^{-11}$ & 1.50$\cdot 10^{-11}$ & 1.36$\cdot 10^{-11}$ & 1.29$\cdot 10^{-11}$ & 1.21$\cdot 10^{-11}$& 1.15$\cdot 10^{-11}$ & 1.07$\cdot 10^{-11}$\\\hline\hline\\[-2mm]
\multicolumn{9}{c}{Ratio of 0.2\,--\,2.0 keV band flux to LETGS 15 lines$^{a}$ flux ($\pm 0.05$ \AA~bands)}\\\hline\\[-3mm]
T (MK) & 0.8 &  1.0 & 1.2 &1.4 & 1.6 & 1.8 & 2.0 & 2.2\\\hline\\[-3mm]
Abus: FIP & - & 1.96 & 2.11 & 2.29 & 2.46 &  2.57 &2.67 & 2.70 \\
Abus: solar &1.64 & 1.65 & 1.75 & 1.76 & 1.92 & 1.96 & 2.00 & 2.04\\\hline
\end{tabular}
\end{center}
{\scriptsize $^{a}$ : \ion{Fe}{xvii} 15.0, \ion{Fe}{xvii} 17.1, \ion{O}{viii} 19.0, \ion{O}{vii} 21.6, \ion{O}{vii} 21.8, \ion{O}{vii} 22.1, 
\ion{N}{vii} 24.8, \ion{C}{vi} 28.5, \ion{N}{vi} 28.8, \ion{N}{vi} 29.5, \ion{C}{vi} 33.7,
\ion{C}{v} 40.3, \ion{C}{v} 41.5, \ion{Si}{xi} 55.3}

\end{table*}

\end{appendix}

\end{document}